\documentclass[%reprint,
twocolumn,
 amsmath,amssymb,
 aps,
prb,
]{revtex4-2}

\usepackage{graphicx}% Include figure files
\usepackage{dcolumn}% Align table columns on decimal point
\usepackage{bm}% bold math
\usepackage{subfigure}
\usepackage{upgreek}

\usepackage{xcolor}
\begin{document}

\preprint{APS/123-QED}

\title{Thermal decomposition of the Kitaev material $\alpha$-RuCl$_3$ and its influence on low-temperature behavior}% Force line breaks with \\

\author{Franziska A. Breitner$^1$}
\email[]{franziska.breitner@physik.uni-augsburg.de}
\author{Anton Jesche$^1$}
 \author{Vladimir Tsurkan$^{2,3}$}
\author{Philipp Gegenwart$^1$}
 \email[]{philipp.gegenwart@physik.uni-augsburg.de}

\affiliation{$^1$Experimental Physics VI, Center for Electronic Correlations and Magnetism, University of Augsburg, Germany}
\affiliation{$^2$Experimental Physics V, Center for Electronic Correlations and Magnetism, University of Augsburg, Germany}
\affiliation{$^3$ Institute of Applied Physics, Chisinau, Moldova}
\date{\today}

\begin{abstract}
%Measurements of the heat capacity and magnetic susceptibility as well as EDX and XRD analysis of heat treated samples of $\alpha$-RuCl$_3$ were performed.
%While EDX reveales an increased ratio of Ru:Cl on the heat treated sample surface, XRD analysis confirmes the formation of RuO$_2$ upon heating the sample to $T\geq 400^\circ$C in Argon atmosphere. The heat treated samples show an increased heat capacity at low temperatures $T<1.5$\,K as well as a decrease in magnetic susceptibility which can consistently be described by the sum of $\alpha$-RuCl$_3$ and a RuO$_2$ fraction which is initially small, but increases with annealing.
%Thermally driven decomposition and reduction of honeycomb ruthenium-trichloride has been known long before the material attracted considerable interest as a Kitaev system.
We explore the effect of heat treatment in argon atmosphere under various  temperatures up to $500^\circ$C on single crystals of $\alpha$-RuCl$_3$ by study of the mass loss, microprobe energy dispersive x-ray spectroscopy, powder x-ray diffraction, electrical resistance as well as low-temperature magnetic susceptibility and specific heat. Clear signatures of dechlorination and oxidation of Ru appear for annealing temperatures beyond $300^\circ$C.
 Analysis of the specific heat below 2~K reveals a RuO$_2$ mass fraction of order $1\%$ for pristine $\alpha$-RuCl$_3$ which increases up to $20\%$ after thermal annealing, fully consistent with mass-loss analysis. The small RuO$_2$ inclusions drastically reduce the global electrical resistance and may thus significantly affect low-temperature thermal transport and Hall effect.

\end{abstract}

\maketitle

%\tableofcontents
%bier hier

\section{\label{}{Introduction}}

The 4d layered spin orbit Mott insulator $\alpha$-RuCl$_3$ \cite{plumb2014alpha, sandilands2016spin, agrestini2017electronically, banerjee2016proximate} is one of the most studied "Kitaev materials" \cite{winter2017,takagi2019}, implying nearest-neighbor bond-directional Ising interactions on the honeycomb lattice~\cite{kitaev2006anyons}. The pure Kitaev model offers an exciting route towards a quantum spin liquid with exotic fractionalized excitations and potential application for topological quantum computation \cite{kitaev2003fault, kitaev2006anyons, hermanns2018physics}. Although $\alpha$-RuCl$_3$ displays a zigzag magnetic order below $T_\mathrm{N}\sim$7-8\,K \cite{sears2015magnetic, sears2017phase, PhysRevB.93.134423}, its intriguing dynamical properties~\cite{banerjee2016proximate,sandilands2015scattering,kim2015kitaev} and the possibility to suppress the order by moderate in-plane magnetic fields~\cite{banerjee2018excitations, baek2017evidence, sears2017phase} led to a strong interest in this material. This was further boosted in 2018 when Kasahara {\it et al.,} reported a half-integer quantized plateau in the thermal Hall conductance, in accordance with chiral Majorana edge modes~\cite{kasahara2018}. Subsequent studies of the thermal Hall effect by different groups however questioned a generic regime of half-quantization and indicated that the thermal Hall conductance is strongly sample dependent
~\cite{yamashita2020,yokoi2021,czajka2021,bruin2022,lefrancois2022,bruin2022origin,czajka2023,zhang2023}.
Oscillatory structures of the magnetothermal conductivity~\cite{czajka2021} were related to coexistent secondary phases that feature differing critical magnetic fields due to stacking disorder~\cite{bruin2022origin}. $\alpha$-RuCl$_3$ has a monoclinic symmetry at room temperature ~\cite{johnson2015monoclinic,PhysRevB.93.134423} and displays a first-order structural transition with large hysteresis around 150 K~\cite{gass2020,kubota2015successive}.
%Indeed it was shown already in 2015 that 
Crystals with structural domains featuring stacking disorder show multiple antiferromagnetic transitions in the specific heat~\cite{johnson2015monoclinic, kubota2015successive, PhysRevB.93.134423}. This holds mainly for powder specimens and low-quality crystals, whose signature in the specific heat is an anomaly near 14~K. High-quality single crystalline samples usually show only one transition at 7\,K, which can be observed in the heat capacity as a sharp peak \cite{banerjee2017neutron}. Stacking disorder can however easily arise in the van der Waals material 
$\alpha$-RuCl$_3$ by non-careful handling or small strain effects during cooling.

In addition, it has been known since 1968 that transition-metal chloride hydrates are chemically unstable and decompose upon heating above 150$^\circ$C~\cite{Newkirk1968}. This raises the question whether a possible degradation of $\alpha$-RuCl$_3$ single crystals may influence its low-temperature physical properties. In particular, if $\alpha$-RuCl$_3$ undergoes a thermally activated degradation, then it could be assumed that already during growth some small fraction of the crystals become degraded. This motivates our systematic study of the effect of moderate temperature treatments on high-quality $\alpha$-RuCl$_3$ single crystals.

%In the region of applied fields of 7.5-11\,T, a magnetically disordered state was found and proposed as field-induced Kitaev QSL \cite{wang2017magnetic,balz2019finite}.
%
In this paper, we report thermal annealing (in Argon atmosphere) studies on $\alpha$-RuCl$_3$ single crystals at temperatures up to 500$^\circ$C. Analysis of the mass loss in combination with EDX and XRD reveals clear evidence for dechlorination and the formation of RuO$_2$ clusters penetrating from the surface into the bulk. While RuO$_2$ inclusions have little influence on magnetic susceptibility as well as on the specific heat anomaly at $T_N$, they dominate over the gapped magnon contribution in $C(T)$ below 2~K. The low-$T$ specific heat reveals approximately 1\% RuO$_2$ even in untreated $\alpha$-RuCl$_3$ single crystals. Our study shows that the bulk electrical conductance is strongly enhanced by metallic RuO$_2$ inclusions suggesting that the latter may also affect the low-$T$ thermal transport properties.

\section{\label{}{Methods}}

High quality crystals of $\alpha$-RuCl$_3$ were grown using vacuum sublimation as described in \cite{reschke2018sub}. Zero-field heat capacity measurements were done to check the quality of all crystals before heat treatments. Thereby, a single transition at 7\,K and no signature at 14\,K were detected. 

\begin{figure}[h]
\includegraphics[width=0.45\textwidth]{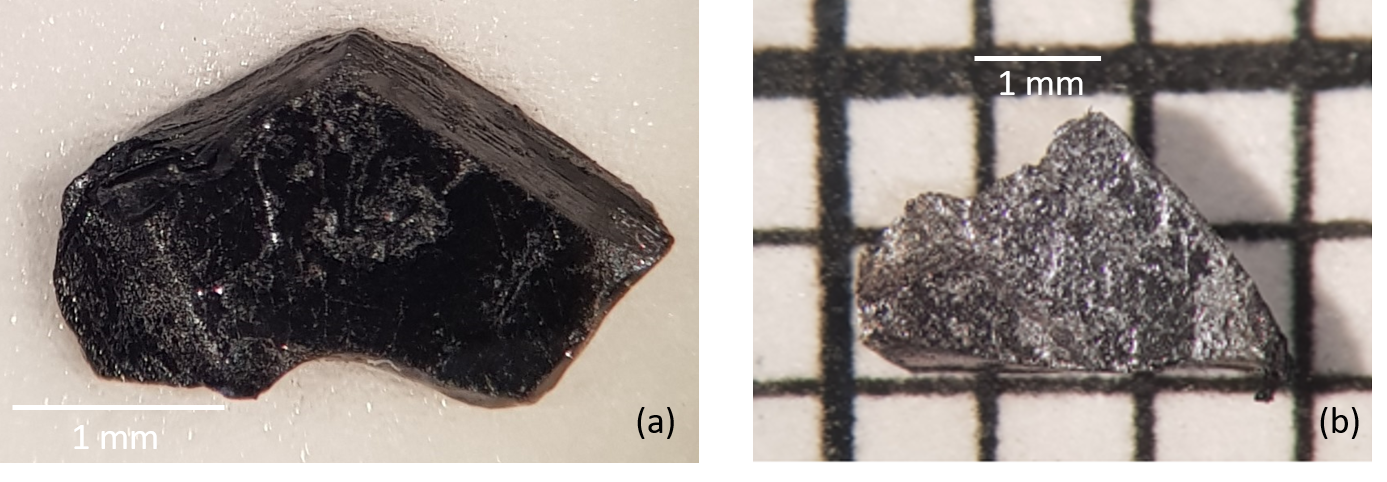}
\caption{\label{crystals} Crystals used for heat capacity study. Sample 1 (a) is shown before heat treatments, sample 2 (b) after being heat treated at 500$^\circ$C in Argon flow.}
\end{figure}

Heat treatments were performed in two different ways, see supplemental material (SM) for a table with all studied samples~\cite{SM}. Two of the crystals (sample 1 and sample 5) were sealed in a quartz ampoule under 150\,mbar Ar atmosphere, after evacuating the tube several times down to $2 \cdot 10^{-2}$\,mbar and flushing with Ar gas, then heated in a muffle furnace up to 400 or 450$^\circ$C for 12\,h. To avoid contamination of the sample the quartz tube was previously cleaned using acetone and then baked out at 70$^\circ$C for one hour before inserting the crystal. The other samples were heat treated using an Al$_2$O$_3$ crucible placed inside a DTA chamber, which was then evacuated down to 3\,mbar and flooded with Argon gas before heating the sample in Ar flow to 500$^\circ$C for 1\,h.

Heat capacity (HC) measurements in the range of 0.35 - 20\,K were performed in a Quantum Design PPMS with Helium-3 Option. The samples were mounted onto the platform using Apiezon N grease. Magnetic susceptibility in the range of 2 - 300\,K was measured utilizing the Quantum Design MPMS 3. The sample was mounted onto a quartz rod using GE varnish and later removed using isopropanol. Electrical transport measurements in the range of 125 - 300\,K were performed in the PPMS utilizing the ETO option. Contacts for four-wire measurements were made using two-component silver epoxy. 

Powder X-ray diffraction measurements were performed using a Rigaku Miniflex600 powder diffractometer (Cu-K$_\alpha$ radiation). A scanning electron microscope (SEM, Merlin Gemini 2, Zeiss) equipped with an energy dispersive x-ray (EDX) analysis probe (X-Max 80N SDD detector, Oxford Instruments) was utilized for structural and compositional investigation. Silver epoxy was used to mount the crystals onto the sample holder.

After each measurement the samples were carefully cleaned using n-butyl acetate to avoid carrying any epoxy or grease residue into the next measurement while at the same time avoiding damage to the crystals.

\section{\label{}{Results and Discussion}}
%consistency check
Before performing any kind of heat treatments we checked whether the specific heat is affected by multiple removals from HC and EDX pucks as it is known that less careful handling can potentially induce stacking disorder that profoundly changes the $T_{\rm N}$ and the signature of magnetic order~\cite{PhysRevB.93.134423}. 
As shown in SM~\cite{SM}, no change in the HC was found, confirming that any changes in our study are induced by heat treatments.

\begin{figure}
\includegraphics[width=0.45\textwidth]{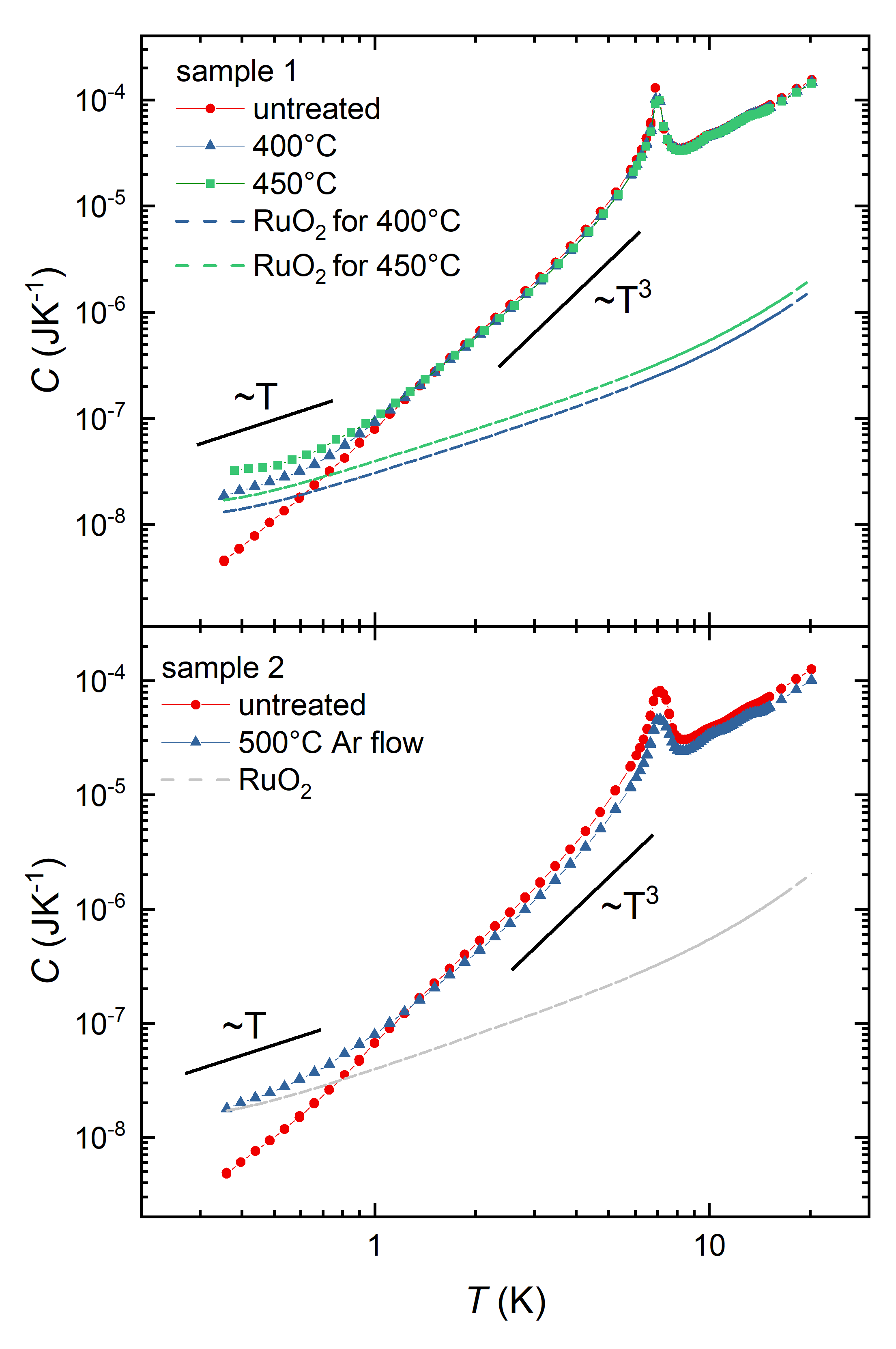}
\caption{\label{HC_temp} Heat capacity of initial samples 1 (a) and 2 (b) compared with those after several heat treatments. For $T<1.5$\,K the heat capacity increases towards low temperatures with increasing annealing temperature, while the peak at 7\,K is reduced by magnitude. For comparison, the fraction of heat capacity attributed to RuO$_2$ is shown as dotted lines.}
\end{figure}

%hc sample 1
For sample 1, heat treatments were performed at increasing temperatures, until a change in the HC could be detected. The first change was observed after heat treatment at 400$^\circ$C. Already an increase of the HC towards low temperatures for temperatures below 1.5\,K as well as a shrinking of the peak at 7\,K can be detected, as can be seen in Fig.~\ref{HC_temp}(a). The procedure was repeated with a maximum heat treatment temperature of 450$^\circ$C. Again, the HC showed an even more pronounced increase towards low temperatures.
Here, the exponential impact of the maximum temperature on the activation process exceeds that of longer dwell time, thus no experiments with longer dwell times were conducted.

%mass loss
After each step the sample mass was determined. Each heat treatment led to a notable decrease, the exact values of which are listed in Tab. \ref{mass}.

%hc sample 2

The heat treatment for sample 2 was performed at 500$^\circ$C in Argon flow. Comparing the HC of the heat treated and the untreated crystal, as is shown in Fig.~\ref{HC_temp}(b), the same increase towards low temperatures and shrinking of the 7\,K peak can be observed.

%mass loss sample 2
Compared to its inital mass of 5.44\,mg the mass of the heat treated sample 2 was reduced to 4.58\,mg indicating a relative mass loss of 16\%. As the sample surface appeared rather porous after heat treatment, very careful handling was required to avoid parts breaking off during transport or handling.

\begin{figure}
\includegraphics[width=0.45\textwidth]{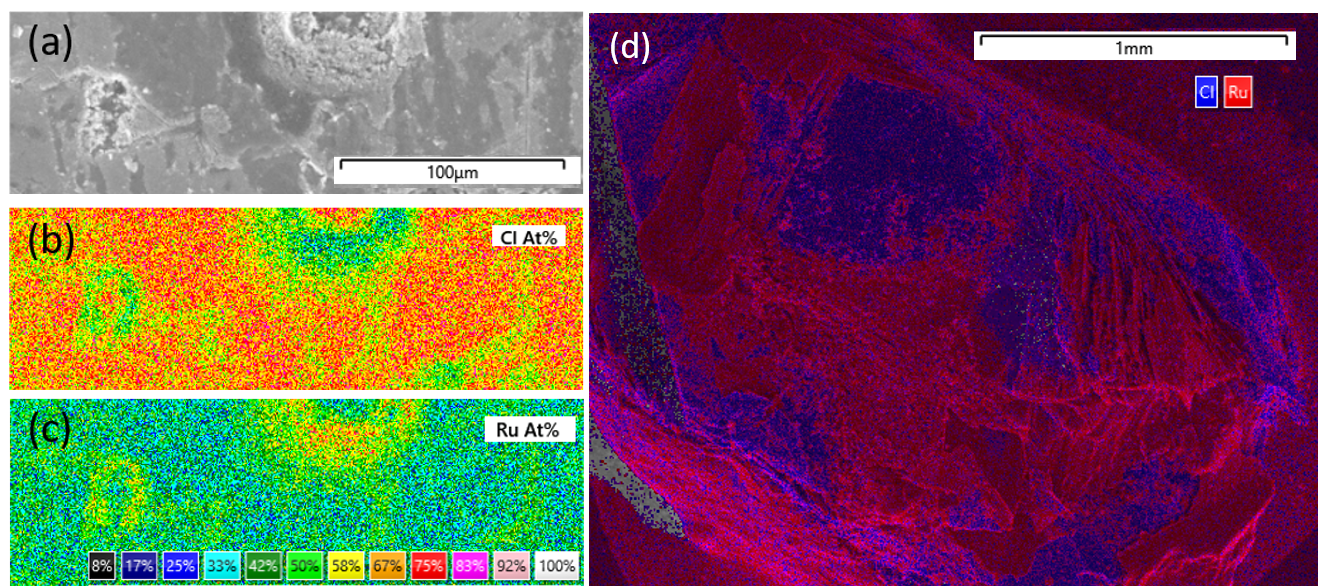}
\caption{\label{EDX_1und2}Elemental maps of Ru and Cl for sample 1 after heat treatment at 400$^\circ$C (a-c) and sample 3 which was treated analogous to sample 2 (d). For sample 1 in some areas a significant increase in the Ru concentration along with a corresponding decrease in the Cl concentration can be observed (e.g. green areas in (b)). Sample 2 displays an overall diminished Cl concentration on the surface, with large parts of the surface showing a majority of Ru.}
\end{figure}

%EDX
EDX was used to determine the sample stoichiometry and map the elemental distribution on the surface. After the initial consistency check, EDX analysis was only performed once the HC changed in order to minimize stress on the crystal. For the untreated crystals, the obtained molar ratio of Ru:Cl amounted to 25(3):75(3), both Ru and Cl were evenly distributed across the observed surface, as can be seen in SM~\cite{SM}. No significant change in elemental distribution was observed for temperatures up to 300$^\circ$C. However, it should be noted, that only a fraction of the crystal surface was evaluated in greater detail due to spatial limitations and time constraints.
Upon heating sample 1 to 400$^\circ$C, the formation of clusters, some as large as 70\,$\upmu$m in diameter, was observed (see Fig.~\ref{EDX_1und2}(a)-(c)). Stoichiometry analysis of such clusters shows a decrease of Cl concentration down to 25\,at\% in some areas and a corresponding increase in Ru concentration. 
%EDX 450
%Looking at the sample after heat treatment at 450$^\circ$C, it proved challenging to analyze the same spot as before in order to allow for direct comparison, as the crystal surface appeared changed due to the formation of new clusters. Overall, clusters similar to those previously found were clearly visible, however, it is difficult to quantify from our analysis whether and if so by what percentage the amount and size of clusters has increased after the second heat treatment. 
Averaging over the investigated surface, the molar ratio of Ru:Cl is determined to be about 30:70 for sample 1 treated at 400$^\circ$C and 32:68 for 450$^\circ$C. We therefore conclude that further degradation of the sample has occurred due to the second heat treatment.

%sample 2
After heat treatment at 500$^\circ$C sample 2 did not show any visible formation of clusters, however the Cl concentration was significantly diminished across the whole crystal surface. The average molar ratio Ru:Cl was determined to be 62:38. 

\begin{figure}
\includegraphics[width=0.45\textwidth]{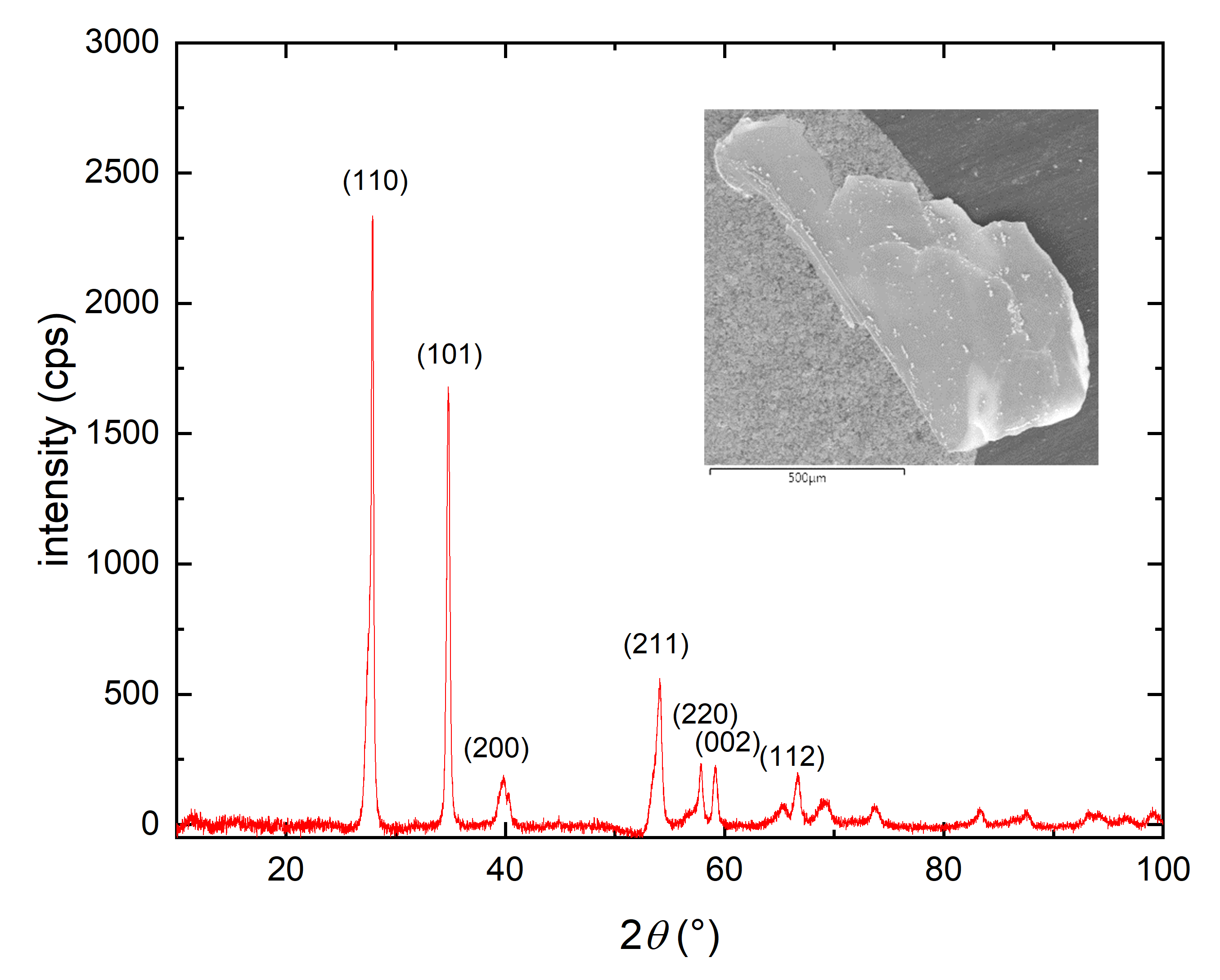}
\caption{\label{XRD} XRD pattern of ground surface material. The peaks match those for RuO$_2$. }
\end{figure}

As EDX analysis is limited to the surface layers due to a penetration depth of the electron beam below $\sim 1\mu$m, the question arises as to how deep into the crystal this effect can still be observed. 
For this purpose, another crystal (sample 3) was prepared as previously described in order to investigate the penetration depth of the degradation process without rendering sample 2 unusable for further measurements. From sample 3 few layers were peeled off, then the crystal was cut into half and EDX was performed on the freshly obtained surfaces.
The distribution of Ru and Cl on the surface can be seen in Fig.~\ref{EDX_1und2}(d), revealing large areas on the crystal surface with predominantly Ru being detected.
While after a few layers, the molar ratio of Ru:Cl is still significantly enhanced to 44:56, roughly halfway into the 0.5\,mm thick crystal only 27\,at\% Ru are detected.
%XRD 
In order to determine whether the accumulated Ru on the surface is metallic ruthenium or some other product, enough material was carefully removed from the crystal surface and ground into fine powder using an agate mortar and pestle in order to perform X-ray powder diffraction. The obtained diffraction pattern shown in Fig.~\ref{XRD} matches that of the metallic transition-metal oxide RuO$_2$~\cite{khachane2008catalytic} while no pure Ru could be detected. 

%Fit HC

\begin{figure} [h]
\includegraphics[width=0.49\textwidth]{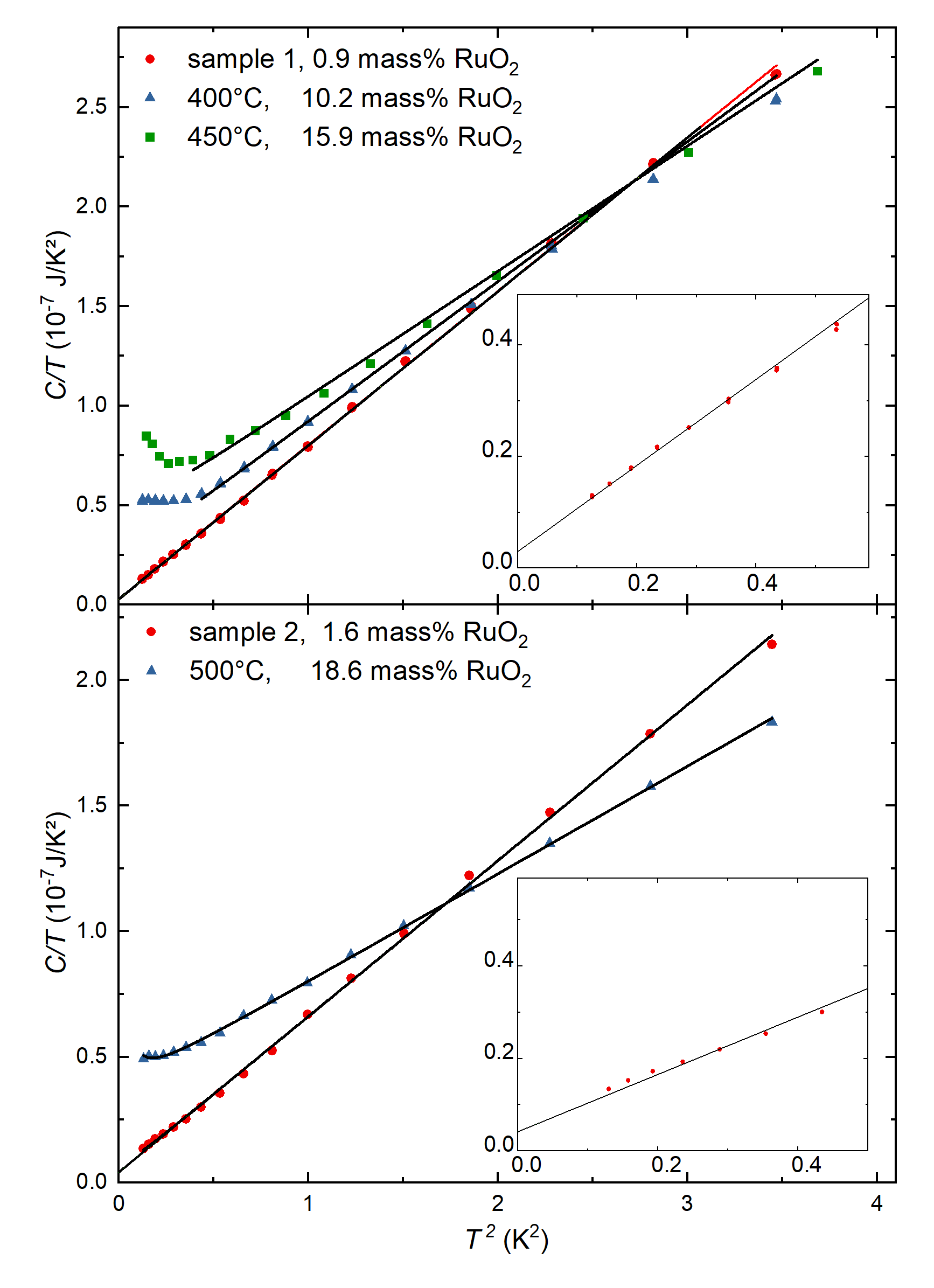}
\caption{\label{CoverT__fit} Heat capacity as $C/T$ vs. $T^2$ for sample 1 (upper) and 2 (lower panel) before (red circles) and after (blue triangles) heat treatments. The black solid lines are fits according $C/T=(m_\mathrm{sample}-m_{\mathrm{RuO}_2})\cdot(\beta_{\mathrm{RuCl}_3}T^2)+m_{\mathrm{RuO}_2}\cdot C_{\mathrm{m,RuO}_2}/T$
with masses as listed in Tab.~\ref{mass}, $\beta_{\mathrm{RuCl}_3}=1.29$ and $1.18\cdot10^{-5}$\,J/gK$^4$ for sample 1 and 2, respectively.
The heat capacity of RuO$_2$ is described by $C_{\mathrm{m,RuO}_2}=M_{\mathrm{RuO}_2}^{-1} (\alpha T^{-2}+\beta T^3 + \gamma T)$ 
with $\alpha = 5.9\cdot 10^{-5}$\,JK/mol, $\beta = 2.25\cdot 10^{-5}$\,J/molK$^4$ and $\gamma = 5.77\cdot 10^{-3}$\,J/molK$^2$~\cite{SM}.
 }
\end{figure}

Using this information we examine the low-$T$ HC of initial and heat-treated $\alpha$-RuCl$_3$ for samples 1 and 2. As shown in Fig.~\ref{CoverT__fit}, the measured data below 1.9 K are described by the sum of two contributions arising from  phonons in $\alpha$-RuCl$_3$ and phonons and electrons in RuO$_2$. Note, that the magnon contribution $\sim \exp(-\Delta/k_BT)$ with $\Delta=1.7$~meV~\cite{banerjee2018excitations} is negligible compared to phonons in this temperature range. For the fit, we used the measured total sample mass and described the total HC (in units of J/K) by the function $C/T=(m_\mathrm{sample}-m_{\mathrm{RuO}_2})\cdot(\beta_{\mathrm{RuCl}_3}T^2)+m_{\mathrm{RuO}_2}\cdot C_{\mathrm{m,RuO}_2}$ with $m_{\mathrm{RuO}_2,\mathrm{HC}}$ as free fit parameter.  The (molar) specific heat of RuO$_2$ was measured on a pellet and found in good agreement to literature~\cite{passenheim1969heat}.   For details, we refer to SM~\cite{SM}. The converted (mass) specific heat $C_{\mathrm{m,RuO}_2}$ was then used in the above fit of the HC. The fit also includes the phonon contribution of $\beta_{\mathrm{RuCl}_3}$, which was determined by fitting the untreated crystals (yielding the parameters given in the caption of Fig.~\ref{CoverT__fit}) and then fixed for all further fits. 
% In order to obtain the mass distribution of both in the crystals, we applied a fit function combining the heat capacities in units of J/gK scaled by the mass, as shown in Fig.\ref{CoverT__fit}. For the mass heat capacity of RuO$_2$ (labeled $C_{\mathrm{m,RuO}_2}$) the data shown in Fig.\ref{RuO2_fit_mol} was converted into units of J/gK. The obtained parameters were $\alpha= 4.43\cdot 10^{-7}$\,JK/g, $\beta= 1.69\cdot 10 ^{-7}$\,J/gK$^4$, $\gamma = 4.34\cdot 10^{-5}$\,J/gK. For RuCl$_3$ we used a $C=\beta_{\mathrm{RuCl}_3}\cdot T^3$ ansatz assuming $\gamma_{\mathrm{RuCl}_3}\sim 0$. As the total mass is known, $m_{\mathrm{RuCl}_3}$ was expressed as the difference of sample mass and RuO$_2$. The remaining fit parameters are $\beta_{\mathrm{RuCl}_3}$ and $m_{\mathrm{RuO}_2}$ the values of which are listed in Tab.\ref{mass}. $\beta_{\mathrm{RuCl}_3}$ was determined from fitting the untreated samples and then fixed for all further fits. For sample 1 we found $\beta_{\mathrm{RuCl}_3}=1.29\cdot10^{-5}$\,J/gK$^4$ and for sample 2 $\beta_{\mathrm{RuCl}_3}=1.18\cdot10^{-5}$\,J/gK$^4$. The fit function can then be written as $C/T=(m_\mathrm{sample}-m_{\mathrm{RuO}_2})\cdot(\beta_{\mathrm{RuCl}_3}T^2)+m_{\mathrm{RuO}_2}\cdot C_{\mathrm{m,RuO}_2}$. The masses resulting from this fit (listed in Tab. \ref{mass})
The fitted values for $m_{\mathrm{RuO}_2,\mathrm{HC}}$ listed in 
Tab.~\ref{mass} are in good agreement with those obtained from the analysis of the weight loss  according to $m_{\mathrm{RuO}_2,\mathrm{scale}}=\Delta m (1-\frac{M_{\mathrm{mol,RuCl}_3}}{M_{\mathrm{mol,RuO}_2}})^{-1}$, where $\Delta m<0$ denotes the measured mass difference between heat-treated and pristine samples, arising by the loss of chlorine and gain of oxygen according to 2RuCl$_3$ + 2O$_2 \rightarrow 2$RuO$_2$ + 3Cl$_2$.
Applying the same fitting procedure to the HC of the pristine samples yields RuO$_2$ masses corresponding to 1-2$\%$ of total sample mass. This suggests that even in unannealed crystals of RuCl$_3$ a tiny RuO$_2$ metal fraction cannot be excluded. Furthermore, the Sommerfeld coefficient of the respective RuO$_2$ contribution can be accessed directly by looking at the intersection of the fit function with the $C/T$ axis, cf. the insets of Fig.~\ref{CoverT__fit}.

%Ru fit
Fitting with the heat capacity with Ru instead of RuO$_2$ yields fits of lower quality with Ru masses significantly higher than what would be possible due to the measured mass loss. 
Another possibility would be to fit the data with a combination of Ru and RuO$_2$. However, fitting with the masses as free parameters results in the same values as obtained for the fit with just RuO$_2$ and the mass of Ru chosen as zero. We therefore conclude that most if not all of the degraded $\alpha$-RuCl$_3$ turns into RuO$_2$ upon heating.

\begin{table}

\begin{ruledtabular}
\begin{tabular}{lcccc}

& $m_\mathrm{sample}$ & $m_{\mathrm{RuO_2, balance}}$  &  $m_{\mathrm{RuO_2, HC}}$ & $\frac{m_\mathrm{RuO_2, HC}}{m_\mathrm{sample}}$ \\
\hline
\hline
\textbf{sample 1} &6.34 mg  & - & 0.06 mg & 0.9\%\\
\hline
400$^\circ$C & 5.96 mg &  0.68 mg & 0.61 mg & 10.2\% \\
\hline
450$^\circ$C & 5.84 mg & 0.89 mg & 0.93 mg & 15.9\%\\
\hline
\textbf{sample 2} & 5.44 mg & - & 0.09 mg & 1.6\% \\
\hline
500$^\circ$C & 4.58 mg & 0.89 mg & 0.85 mg & 18.6\%  \\

\end{tabular}
\end{ruledtabular}
\caption{\label{mass}Comparison of the masses of RuO$_2$ in the initial and heat treated samples determined via fit to the low-$T$ heat capacity versus values calculated from mass loss determined by balance.}
\end{table}

%MPMS
Another crystal (sample 4) was used to investigate the change in magnetic behavior due to heat treatment. The sample was tempered at 500$^\circ$C in argon flow for 1\,h analogous to sample 2. The susceptibility measurement of the heat treated crystal showed significantly lower absolute values compared to the initial measurement. Scaling with the $\alpha$-RuCl$_3$ and RuO$_2$ concentration determined from mass loss resulted in a plot showing good agreement with the first measurement.

\begin{figure}
\includegraphics[width=0.45\textwidth]{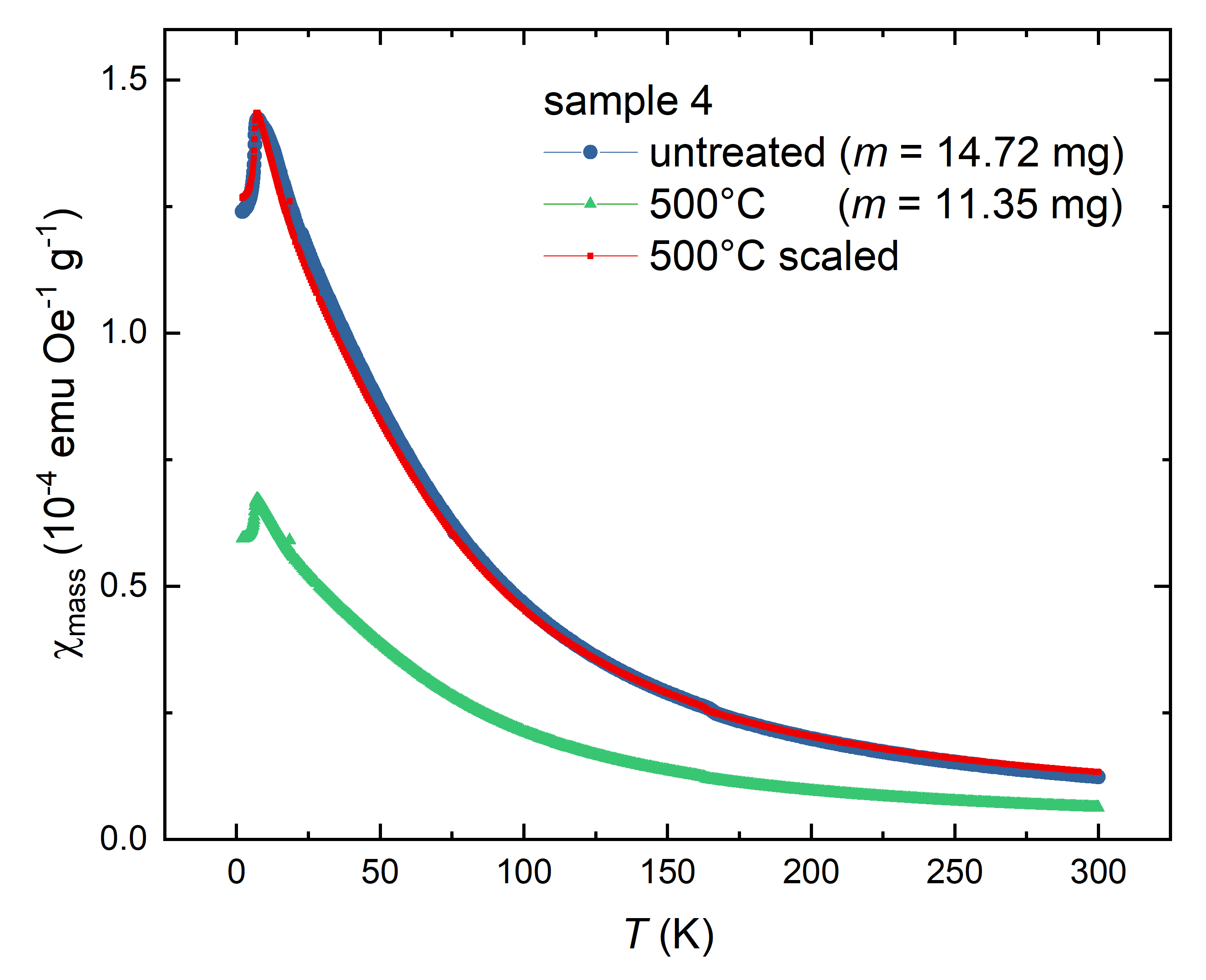}
\caption{\label{MPMS} Temperature dependence of the magnetic susceptibility measured with a magnetic field of $H=1\,\mathrm{T}$ applied along the $ab$-plane before and after heat treatment at 500$^\circ$C. The values given in brackets refer to the sample masses, wherein the heat treated sample contains 6\,mg RuO$_2$ according to calculation from mass loss. Scaling the heat treated measurement (green) with a corrected mass of $\alpha$-RuCl$_3$ due to degradation yields a plot (red) corresponding to that of the initial sample (blue).} 
\end{figure}

%HC fit aus appendix

%rho

Electrical transport measurements were performed on a pristine and 400$^\circ$\,C heat treated crystal (sample 5). While the untreated sample clearly shows insulating behaviour, the resistivity of the heat treated sample changes by several orders of magnitude. From previous analysis, we know that after heat treatment at 400$^\circ$C only $\sim$10\% of the sample consist of RuO$_2$ resulting in a deviation visible only at very low temperatures in heat capacity, yet the influence on the electronic transport properties is significant over the whole temperature range. For comparison we also plotted the resistivity of pure RuO$_2$ \cite{butler1971crystal}, revealing that the values of the heat treated sample are already closer to that of RuO$_2$ than $\alpha$-RuCl$_3$. Since milli-Kelvin thermal conductance of metallic RuO$_2$ is far higher than that of insulating $\alpha$-RuCl$_3$, we expect a significant influence of RuO$_2$ inclusions on the low-$T$ thermal transport and Hall effect.

\begin{figure}
\includegraphics[width=0.45\textwidth]{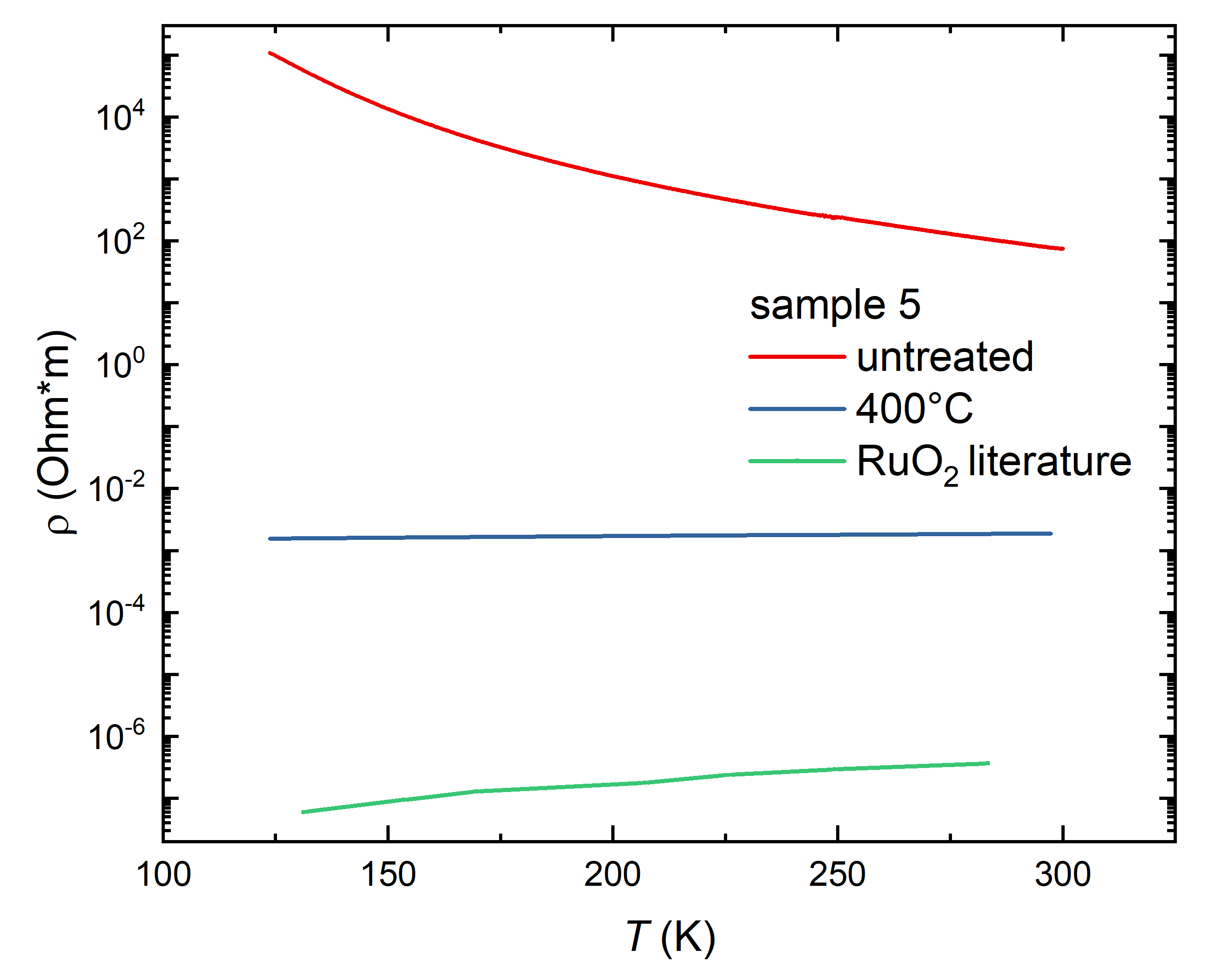}
\caption{\label{rho} Measurement of the electrical transport in the temperature range of 125-300\,K for an untreated (red) and heat treated at 400$^\circ$\,C (blue) $\alpha$-RuCl$_3$ crystal. The green line shows literature data \cite{butler1971crystal} of pure RuO$_2$.} 
\end{figure}

%oxygen?
Remarkably we find a complete oxidation of dechlorinated ruthenium in our experiments, despite sealing the samples in an ampoule which was evacuated and flushed with Argon gas several times. In air $\alpha$-RuCl$_3$ is extremely sensitive to decomposition and oxidation already under very moderate heating~\cite{Newkirk1968}. While it is well established that $\alpha$-RuCl$_3$ needs to be handled mechanically with maximal care to avoid the formation of stacking faults, our experiments indicate that in addition special care is needed to avoid degradation and oxidation. This concerns for instance long-term storage in air or baking of glued metal wire contacts on the crystal surface, required for thermal transport measurements. Protected gas atmosphere is recommended, along with careful check for partial dechlorination and oxidation. 

It should further be noted, that XRD analysis of a different batch revealed the as purchased powder to not be pure $\alpha$-RuCl$_3$, but rather containing some RuO$_2$ and other impurities. This would offer a possible explanation for the small (of order ~1\%) RuO$_2$ fraction in the initial crystals, yet not for the increase in RuO$_2$ for annealed crystals.
While the exact origin of oxygen for the degradation reaction in our study remains unclear~\cite{SM}, the above mentioned observations along with the heat capacity data lead us to conclude that even for unannealed crystals the presence of a small percentage of metallic RuO$_2$ cannot be excluded.

\section{\label{}{Conclusion}}
In conclusion, we performed heat treatments on $\alpha$-RuCl$_3$ single crystals in closed Argon atmosphere up to 450$^\circ$C as well as Argon flow up to 500$^\circ$C in order to investigate the impact of annealing on the low temperature physical properties. Both samples showe enhanced heat capacity towards low temperatures for T$<$1.5\,K. SEM and EDX revealed the formation of Ru rich clusters on the surface of sample 1 and an overall decreased percentage of Cl on the surface of sample 2, after the samples were heated to at least 400$^\circ$C. 
Dechlorination and oxidation takes place beneath the sample surface to some degree, however seems to be less pronounced towards the center of the investigated crystal. Powder XRD analysis revealed the surface material to be RuO$_2$. The deviation of both heat capacity and susceptibility measured after heat treatment from the initial measurement can be explained by a decreased amount of $\alpha$-RuCl$_3$ along with the formation of RuO$_2$. The RuO$_2$ content determined via fit in both cases agrees well with the value calculated from the measured mass loss and amounts to 10-20 mass$\%$. Such relatively small mass fraction of metallic RuO$_2$ already reduces the electrical resistance of degraded $\alpha$-RuCl$_3$ by several orders of magnitude. Importantly, the low-temperature specific heat analysis of pristine $\alpha$-RuCl$_3$ crystals (before thermal treatment) also yields the presence of 0.9-1.6 mass$\%$ RuO$_2$. It would be important to clarify whether such a low fraction of metallic inclusions as found in pristine crystals, though effectively invisible in most physical properties, may have an impact on the low-temperature thermal transport and Hall effect in $\alpha$-RuCl$_3$ crystals, as found in our electrical transport measurements.
\\

\section*{Acknowledgements}
We are grateful to Alexander Herrnberger and Klaus Wiedenmann
for technical support and acknowledge fruitful discussions with Alexander A. Tsirlin, A. Loidl, Y.-J. Kim, S.E. Nagler and R. Valenti. This work was supported by the German Science Foundation through TRR80 (Project No. 107745057). Partial support of ANCD via project 20.80009.5007.19 is acknowledged.

\bibliography{bibpaperRuCl3}

\end{document}